\documentclass[a4paper,fleqn,usenatbib]{mnras}
\usepackage[T1]{fontenc}
\usepackage{ae,aecompl}
\usepackage{graphicx}
\usepackage{amsmath}
\usepackage{amssymb}
\usepackage{txfonts}

\usepackage{float}
\usepackage{adjustbox}

\newcommand{\ltsima}{$\; \buildrel < \over \sim \;$}
\newcommand{\lsim}{\lower.5ex\hbox{\ltsima}}
\newcommand{\gtsima}{$\; \buildrel > \over \sim \;$}
\newcommand{\gsim}{\lower.5ex\hbox{\gtsima}}
\newcommand{\bra}{\langle}
\newcommand{\ket}{\rangle}

\newcommand{\dd}{\mathrm{d}}

\title[Information entropy in cosmology]
{Information entropy in cosmological inference problems}
\author[A.M.M. Pinho, R. Reischke, M. Teich, B.M. Sch{\"a}fer]
{Ana Marta Pinho$^1$, Robert Reischke$^{2,3,4}$, Marie Teich$^4$, Bj{\"o}rn Malte Sch{\"a}fer$^4$\thanks{e-mail: bjoern.malte.schaefer@uni-heidelberg.de}\\
$^1$Institut f{\"u}r theoretische Physik, Universit{\"a}t Heidelberg, Philosophenweg 16, 69120 Heidelberg, Germany\\
$^2$ Department of Physics, Technion, Haifa 32000, Israel
\\
$^3$ Department of Natural Sciences, The Open University of Israel, 1 University Road, P.O. Box 808, Ra'anana 4353701, Israel\\
$^4$Zentrum f{\"u}r Astronomie der Universit{\"a}t Heidelberg, Astronomisches Rechen-Institut, Philosophenweg 12, 69120 Heidelberg, Germany
}

\begin{document}
\pagerange{\pageref{firstpage}--\pageref{lastpage}}
\pubyear{2020}
\maketitle
\label{firstpage}

\begin{abstract}
The subject of this paper is a quantification of the information content of cosmological probes of the large-scale structures, specifically of temperature and polarisation anisotropies in the cosmic microwave background, CMB-lensing, weak cosmic shear and galaxy clustering, in terms of Information theory measures like information entropies. We aim to establish relationships for Gaussian likelihoods, between conventional measures of statistical uncertainties and information entropies. Furthermore, we extend these studies to the computation of (Bayesian) evidences and the power of measurement to distinguish between competing models. We investigate in detail how cosmological data decreases information entropy by reducing statistical errors and by breaking degeneracies. In addition, we work out how tensions between data sets increase information entropy and quantify this effect in three examples: the discrepancy in $\Omega_m$ and $\sigma_8$ between the CMB and weak lensing, the role of intrinsic alignments in weak lensing data when attempting the dark energy equation of state parameters, and the famous $H_0$-tension between Cepheids in the Hubble keystone project and the cosmic microwave background as observed by Planck.
\end{abstract}


\section{introduction}
At the moment, we are observing a natural progression in cosmological data analysis: Firstly, the homogeneous and isotropic background expansion of the Universe was probed with Cepheid variable stars and with supernovae of type Ia, secondly, the linear perturbations in the metric were observed with the temperature and polarisation anisotropies in the cosmic microwave background. Now, and thirdly, the nonlinearly evolved cosmic large-scale structure is dissected by galaxy clustering and weak lensing surveys, where a number of complications arise in data analysis, related to systematic astrophysical effects on one side and to non-Gaussian statistics on the other. With these observations it is possible to investigate the expansion dynamics of the Universe with the relevant laws of gravity and the properties of cosmological fluids at high precision, as well as the initial conditions of structure formation and the processes that lead to the cosmic structures that we see today. In the spirit of narrowing down the allowed parameter range, the combination of cosmological probes is of particular importance, because they are sensitive to different signatures of the cosmological model. While the statistical error is expected to decrease if the probes are consistent, any tension between the best fit-values can hint, if significant, at the presence of systematic errors due to badly understood astrophysical processes, or better, at new physics beyond $\Lambda$CDM or $w$CDM.

The knowledge on cosmological models and their corresponding parameter choices is encapsulated in the likelihood function, which assumes an approximate Gaussian shape if the model has a low complexity and if the data is well-constraining the parameter space \citep[][the latter for an application in cosmology]{fisher_logic_1935, trotta_bayesian_2017}. In this case, nonlinearities in the model can be approximated well enough by linear relationships, which renders the likelihood ideally Gaussian and makes it accessible to the Fisher-matrix formalism \citep[with an application to cosmology,][]{wolz_validity_2012, crittenden_fables_2012, elsner_fast_2012, khedekar_cosmology_2013}, which is ubiquitous in modern cosmology \citep{tegmark_karhunen-loeve_1997, coe_fisher_2009, bassett_fisher4cast_2009, schafer_describing_2016}. Under exactly the assumption of a Gaussian likelihood, it is possible to compute the expected parameter covariance from the second derivatives of the logarithmic likelihood, which becomes equal to the expectation value of the product of first derivatives if averaged over the expected data. 

The Fisher-matrix formalism is foremost a tool for determining statistical errors for a Gaussian likelihood, where in forecasting applications the true model is already known, which in this context is referred to as the fiducial cosmology. 
Because inference from cosmological data is often not limited by statistics but rather by systematics, extensions to the Fisher-formalism have been introduced that allow the forecasting of systematical errors, i.e. the shift of the best-fit point of a Gaussian likelihood if an unknown systematic is not removed or properly modelled \citep{loverde_magnification-temperature_2007, taburet_biases_2009, amara_systematic_2008, schaefer_implications_2009, schafer_parameter_2010, kirk_optimising_2011}.

There is no unambiguous way to quantify the total statistical error budget of a cosmological probe or the significance of tensions between likelihood obtained with different cosmological probes, even in the case of Gaussian likelihoods. Such a measure of total error would be convenient in quantifying the information content of a particular cosmological probe, or its parameter degeneracy breaking power, or in applications of experimental design, where one optimises a survey to yield the smallest possible errors \citep{jenkins_power_2011, kerscher_model_2019}. In a larger context, Bayesian evidences \citep{trotta_applications_2007,trotta_bayes_2008, santos_bayesian_2016} used in model selection are measures of the consistency between likelihood and prior \citep{liddle_model_2006}, and are only one possible choice among many others, for instance the Akaike information or the Bayes-information, for preferring a particular model, incorporating a tradeoff between the goodness-of-fit and model complexity \citep{liddle_present_2006, mukherjee_model_2006, heavens_model_2007, knuth_bayesian_2015}. 

The likelihood $\mathcal{L}(\boldsymbol{D}|\boldsymbol{x},M)$ of a cosmological model $M$, subject to the parameters $\boldsymbol{x}$, in the light of the data set, which we compress into a data vector $\boldsymbol{D}$,  is embedded into Bayes' theorem \citep[for a summary of applications of Bayes-statistics in cosmology, see][]{loredo_bayesian_2012}. 
This expresses the state of knowledge after carrying out an experiment, the posterior $p(\boldsymbol{x}|M,\boldsymbol{D})$, as proportional to the likelihood provided by the experiment times the prior distribution $\pi(\boldsymbol{x}|M)$,
\begin{equation}
p(\boldsymbol{x}|M,\boldsymbol{D}) = \frac{\mathcal{L}(\boldsymbol{D}|\boldsymbol{x},M)\pi(\boldsymbol{x}|M)}{p(\boldsymbol{D}|M)},
\end{equation}
where the constant of proportionality is the inverse of the evidence $p(\boldsymbol{D}|M)$ for the model $M$,
\begin{equation}
p(\boldsymbol{D}|M) = \int\dd^n x\:\mathcal{L}(\boldsymbol{D}|\boldsymbol{x},M) \pi(\boldsymbol{x}|M).
\end{equation}
Note that we compress the data and the parameters into vectors $\boldsymbol{D}$ and $\boldsymbol{x}$ of dimension $m$ and $n$ respectively  We write vector components as $x^\mu$ and dual vector components as $x_\mu$.
In cosmology, one often works under the assumption of Gaussian distributions, where every result can be expressed in terms of the covariance matrix, if the Gauss-Markov-theorem is fulfilled. Models can be well constrained by data if their likelihoods are peaked and if their parameter covariances assume small numbers: In this case, the nonlinearities of the model, which give rise to non-Gaussian likelihoods, can be linearised.

In many studies, the focus is on the statistical errors of cosmological parameters as way to understand whether future data can help to investigate new models for gravity, dark energy or inflationary structure formation. When differentiating between models, for instance between dark energy and a cosmological constant, one introduces an interpolating parameterisation and is content if that parameter has a sufficiently small error for distinguishing between the models. Additionally, evidence measures take the complexity of the model into account and prefer simpler models unless a more complex model explains data significantly better \citep{handley_quantifying_2019}. As such, Bayesian evidence was employed for selecting the most likely model, irrespective of the specific parameter choice. Motivated by the Neyman-Pearson-lemma (which itself is related to a relative entropy), one compares competing models by constructing their logarithmic evidence ratio \citep{trotta_forecasting_2007} as one would do when comparing likelihoods,
\begin{equation}
\Delta B = \ln\frac{p(\boldsymbol{D}|M_1)}{p(\boldsymbol{D}|M_2)}
\end{equation}
such that positive values of $\Delta B$ prefer the model $M_1$ over $M_2$ and vice versa. Quantitatively, one uses the Jeffrey's scale for preferring one model over another \citep{nesseris_is_2013}. Studies concerning Bayesian evidence, or Akaike- or Bayes-information criteria as precursors, have been used for quantifying how well measurements can differentiate between competing models and for optimising experimental design \citep{mukherjee_model_2006}.

Information entropies, on the other side, quantify the amount of randomness in a distribution. The Shannon-entropy $S$ \citep{Shannon_original} is defined as
\begin{equation}
S = -\int\dd^n x\:p(\boldsymbol{x})\ln p(\boldsymbol{x}),
\label{eqn_shannon}
\end{equation}
would assume small numbers for a peaked likelihood, as $S = \ln[2\pi\sigma^2\exp(1)]/2$ for a Gaussian distribution. More general measures of entropy are R{\'e}nyi-entropies $S_\alpha$ \citep{Renyi_original, RenyiEntGaussProc} that are parameterised by $\alpha>0$ and $\alpha\neq 1$,
\begin{equation}
S_\alpha = -\frac{1}{\alpha-1}\ln\int\dd^n x\:p(\boldsymbol{x})p^{\alpha-1}(\boldsymbol{x}),
\end{equation}
where one recovers the Shannon-entropy in the limit $\alpha\rightarrow 1$ by application of de l'H{\^o}pital's rule. R{\'e}nyi-entropies increase likewise with the variance for positive values of $\alpha$, $S_\alpha = \ln(2\pi\sigma^2\alpha^\frac{1}{\alpha-1})/2$. This implies that entropies provide a way of quantifying how constraining data is as they quantify the size of the allowed parameter space \citep{mehrabi_information_2019}.

The subject of our paper is the application of information entropy to cosmology: Similar to \citet{carron_probe_2011}, we investigate the information content of cosmological probes by computing the entropy of their likelihood, and compute relative entropies if cosmological probes are combined \citep{grandis_information_2016}: In this way, we intend to provide an interpretation of cosmological likelihoods in terms of a quantity that is defined axiomatically, as done by Shannon, without any ambiguity. We will show that even more complex quantities such as biases between likelihoods or Bayesian evidences can be expressed as relative entropies, giving them again an axiomatically defined meaning and a natural scale for their magnitude. It is the case that even the dark energy figure of merit designed for quantifying the performance of cosmological probes to measure deviations from the cosmological constant $\Lambda$ is actually an information entropy. While focusing on Gaussian distributions, where all calculations we show have analytical solutions, the concept of information entropy is perfectly applicable to asymmetric or even multimodal distributions, and provides natural generalisations to quantities that are intuitive for Gaussian distributions. This applies in particular to Bayesian-evidences or evidence ratios, which can in fact be related to information entropy differences, which are interpretable directly without resorting to the rather arbitrarily defined Jeffreys-scale and are likewise quantified in units of nats.

Additionally, the Shannon-entropy singles out the Gaussian distribution as being extremal: Among all distributions with a fixed variance, the Gaussian distribution maximises the Shannon-entropy $S$, which is usually shown by functional variation with a boundary condition. We would like to illustrate this statement using a Gram-Charlier-parameterised distribution $p(x)\dd x$ with weak non-Gaussianities \citep{wallace_asymptotic_1958} described by the cumulants
$\kappa_3$ and $\kappa_4$, both of which are much smaller than one,
\begin{equation}
p(x) = \frac{1}{\sqrt{2\pi\sigma^2}}\exp\left(-\frac{x^2}{2\sigma^2}\right) \Bigg[1+\frac{\kappa_3}{3!\sigma^3}H_3\left(\frac{x}{\sigma}\right)+\frac{\kappa_4}{4!\sigma^4}H_4\left(\frac{x}{\sigma}\right) \Bigg],
\end{equation}
with the Hermite-polynomials $H_n(x)$ of order $n$. Substituting this series into the definition~(\ref{eqn_shannon}) and approximating $\ln(1+\epsilon)\simeq\epsilon$ for $|\epsilon|\ll 1$, one obtains at second order the result
\begin{equation}
S = \frac{1}{2}\ln\left[2\pi\sigma^2\exp(1)\right] - \frac{1}{3!}\frac{\kappa_3^2}{\sigma^6} - \frac{1}{4!}\frac{\kappa_4^2}{\sigma^8},
\label{eqn_entropy_gram_charlier}
\end{equation}
by using the orthogonality relation of the Hermite-polynomials,
\begin{equation}
\int\dd x\:\frac{1}{\sqrt{2\pi\sigma^2}}\exp\left(-\frac{x^2}{2\sigma^2}\right)\:H_m\left(\frac{x}{\sigma}\right)\:H_n\left(\frac{x}{\sigma}\right) = n!\delta_{mn}.
\label{eqn_hermite}
\end{equation}
Eqn.~(\ref{eqn_entropy_gram_charlier}) shows that the entropy of the Gaussian distribution is always diminished by non-Gaussianities, because $\kappa_3^2$ and $\kappa_4^2$ are as squares necessarily positive. Because of this result, we would like to point out that the information entropies that we compute are upper bounds, and that realistic non-Gaussian likelihoods would have lower values for their information entropies than their Gaussian counterparts. It is remarkable that the orthogonality relation~(\ref{eqn_hermite}) cancels the influence of non-Gaussianities on $S$ to first order, which can be shown by substituting $1 = H_0(x)$ and $x^2 = H_2(x) + H_0(x)$. Sadly, there is no analogous result to eqn.~(\ref{eqn_hermite}) for the R{\'e}nyi-entropy, but eqn.~(\ref{eqn_hermite}) can be generalised in principle to hold for non-Gaussianities of arbitrary order $\kappa_n$. While this could serve as an illustration, it is by no means a stringent proof, as the Gram-Charlier-expansion is not necessarily positive for all choices of $\kappa_n$.

A multivariate generalisation of eqn.~(\ref{eqn_entropy_gram_charlier}) can serve as a way to estimate the Shannon-entropy from MCMC-samples of the likelihood without the need of a density estimate as a way to compute $\ln p(x)$. Instead, one would estimate multivariate cumulants from the samples directly and correct the Gaussian result for the information entropy. It can be expected that a similar relationship exists for DALI-approximated likelihoods \citep{Sellentin_2014, Sellentin_2015}.

 In our investigation, we will juxtapose two popular cosmologies, $\Lambda$CDM and $w$CDM, with a prior on spatial flatness, parameterised by $\Omega_m$, $\sigma_8$, $h$, $n_s$, $\Omega_b$ and possibly the equation of state $w$ unequal to zero. The fiducial parameter choices are the values of the \cite{Planck2018} ($TT$, $TE$, $EE$, low$z$ + lensing) $\Omega_m = 0.3153$, $\sigma_8 = 0.8111$, $h = 0.6736$, $n_s = 0.9649$ and $\Omega_b = 0.0493$. The recombination redshift is chosen as $z_{re}= 11.357$ and galaxy bias as $b=0.68$ \citep{ferraro_wise_2015}. The dark energy fluid is described by an equation of state parameter constant in time \citep{ChevallierPolarski2001, Linder2003}, where $w = -1$ recovers the case of a cosmological constant $\Lambda$, which is our fiducial for both cosmological models.

After summarising the key concepts of Bayesian statistics, the Fisher-matrix formalism and information entropy in Sect.~\ref{sect_theory}, we demonstrate the decrease in entropy through combining cosmological probes, outlined in Sect.~\ref{sect_probes} in Sect.~\ref{sect_decrease} as well as their correspondence to more conventional measures of error in Sect.~\ref{sect_statistics}. We consider three topical cases of tensions in Sect.~\ref{sect_systematics} and their corresponding loss in evidence and increase in information entropy in Sect.~\ref{sect_evidence}, before summarising our results in Sect.~\ref{sect_summary}. In our investigation we assume the characterics of the Euclid for the large-scale structure and of Planck for the CMB and CMB-lensing.

\section{statistics in cosmology}\label{sect_theory}
Approximating likelihoods with a multivariate Gaussian distribution is the basis of the Fisher-matrix formalism, because the covariance matrix can be computed from the averaged gradients of the logarithmic likelihood. Fixing the fiducial model, one obtains for the Fisher-matrix components $F^\mu_{\  \nu}$ \citep{tegmark_karhunen-loeve_1997},
\begin{equation}
F^\mu_{\  \nu} = -\left\bra\frac{\partial^2\ln\mathcal{L}}{\partial x_\mu\partial x^\nu}\right\ket,
\end{equation}
yielding for a measurement of multipole moments $A_{\ell m}$ and $B_{\ell m}$ of Gaussian random fields $A(\theta,\varphi)$ and $B(\theta,\varphi)$ that are described by angular spectra $C_{AA}(\ell)$, $C_{BB}(\ell)$ and $C_{AB}(\ell)$ as
\begin{equation}
\label{eq:fisher}
F^\mu_{\  \nu} = \sum_\ell \frac{2\ell+1}{2}\mathrm{tr}\left(\frac{\partial}{\partial x_\mu}\ln \boldsymbol{C}_D\:\frac{\partial}{\partial x^\nu}\ln \boldsymbol{C}_D\right),
\end{equation}
where the spectra are combined into a common data covariance $\boldsymbol{C}_D = \boldsymbol{C}_D(\ell)$.

Quoting the logarithmic curvature $F^\mu_{\  \nu}$ of the likelihood surface in parameter space, the tensor of second moments $\boldsymbol{C} = \bra \boldsymbol{x}\otimes\boldsymbol{x}\ket$ or confidence intervals is equivalent for a Gaussian distribution. As the Fisher-matrix corresponds to the inverse parameter covariance $\boldsymbol{C} = \boldsymbol{F}^{-1}$, one can write down a multivariate Gaussian distribution as
\begin{equation}
p(\boldsymbol{x}) = \sqrt{\frac{\mathrm{det}(\boldsymbol{F})}{(2\pi)^n}}\exp\left(-\frac{1}{2}x^\mu F_{\mu\nu} x^\nu\right),
\end{equation}
if one uses coordinates relative to the best-fit point. Note that $\boldsymbol{C}$ is a tensor constructed on the tangent bundle of the parameter space, unlike in Eq. (\ref{eq:fisher}) where $\boldsymbol{C}_D$ is the covariance of the data. Furthermore we will employ the sum convention so that repeated indices are summed over.

Measures of total uncertainty can be derived from the Fisher-matrix in a straightforward way as, for instance, the invariant trace $\mathrm{tr}(\boldsymbol{F})$, the Frobenius-norm $\mathrm{tr}(\boldsymbol{F}^2)$ or the determinant $\mathrm{det}(\boldsymbol{F})$ of which we will take the logarithm $\ln\mathrm{det}(\boldsymbol{F})$ to make the connection to information entropies clearer. Generalisations to the trace and the Forbenius-norm of the type $\mathrm{tr}(\boldsymbol{F}^p)$ with $p>2$ would be restricted in the values that they can assume by the H{\"o}lder-inequality,
\begin{equation}
\frac{1}{n} \mathrm{tr}(\boldsymbol{F}) \leq \left(\frac{1}{n}\mathrm{tr}(\boldsymbol{F}^p)\right)^{\frac{1}{p}}
\label{eqn_hoelder}
\end{equation}
for arbitrary powers $p$ of Fisher-matrices in $n$ dimensions, where the traces can be generalised to arbitrary real-valued powers $p$ by using $\mathrm{tr}(\boldsymbol{F}^p) = \mathrm{tr}\exp(p\ln(\boldsymbol{F}))$.

On the other side, the generalised inequality of the arithmetic and geometric mean implies
\begin{equation}
\frac{1}{n} \mathrm{tr}(\boldsymbol{F}) \geq \mathrm{det}(\boldsymbol{F})^{\frac{1}{n}},
\label{eqn_arith_geo}
\end{equation}
such that the information entropies are bounded by traces of the Fisher-matrix, as shown in the next paragraph. Specifically, while $\mathrm{tr}(\boldsymbol{F}) = \sum_\mu \sigma_\mu^{-2}$ is a measure of the total uncertainty of the likelihood, it does not differentiate between correlated and uncorrelated distributions, which is taken care of by $\mathrm{tr}(\boldsymbol{F}^2)$, as the expression contains information from the off-diagonal elements in addition to performing a different weighting of the errors. While all trace-relations for arbitrary $p$ are measures of total error, only the determinant provides a geometric interpretation as the volume of parameter space: The dark energy figure of merit is defined as the volume of the $w_0$-$w_a$ subspace of parameter space bounded by the $1\sigma$-contour. Of all these measures, however, only $\mathrm{tr}(\boldsymbol{F})$ is additive for statistically independent measurements. Given the inequalities~\ref{eqn_hoelder} and~\ref{eqn_arith_geo}, we state all results in a scaled way, i.e. $(\mathrm{tr}(\boldsymbol{F}^p)/n)^{1/p}$ and $\mathrm{det}(\boldsymbol{F})^{1/n}$. The usage of these scaled traces is motivated by the fact that for a diagonal Fisher-matrix with identical entries $1/\sigma^2$ they all return the same value of $1/\sigma^2$ irrespective of $n$ or $p$. Lastly, the Fisher matrix can also be understood as a metric tensor in the context of information geometry by describing the parameter space as a Riemaniann manifold as studied in \citet{amari_information_2016} and applied to a cosmological setting in \citet{giesel2020information} yielding insights about the geometrical structure of parameter spaces.

\begin{figure}
\centering
\includegraphics[scale=0.45]{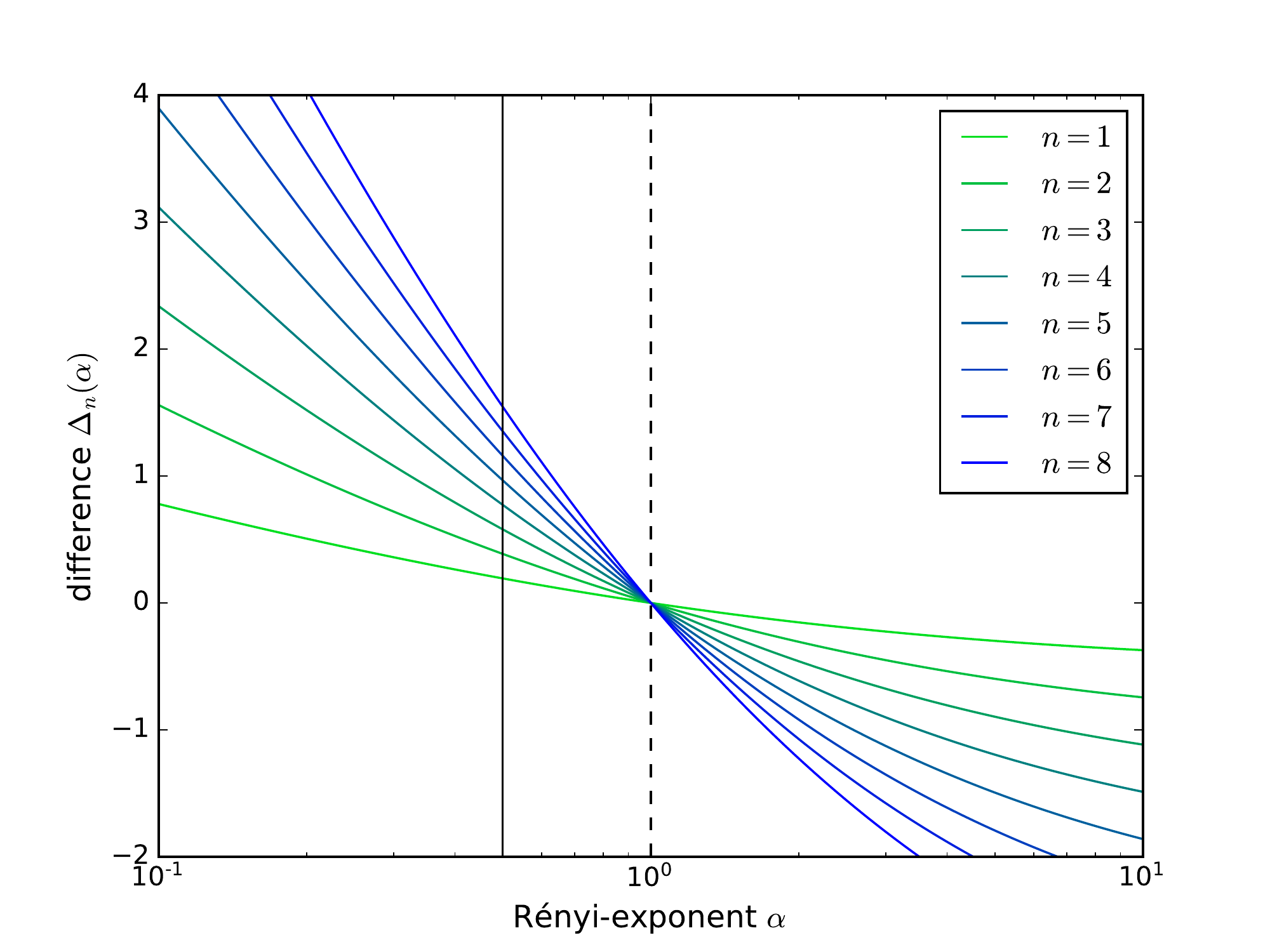}
\caption{Difference $\Delta_n(\alpha)$ between R{\'e}nyi- and Shannon-entropies as a function of $\alpha$ for $n$-variate Gaussian distributions with identical covariances. The dashed vertical line corresponds to the Shannon-case, the solid vertical line to the Bhattacharyya-case with $\alpha=1/2$.}
\label{fig_entropy_scaling}
\end{figure}

Analytical expressions for the entropies $S$ and $S_\alpha$ can be derived for a multivariate Gaussian in terms of the determinant of the covariance matrix $\boldsymbol{C}$ as the inverse Fisher-matrix. Specifically, integration by substitution yields directly
\begin{equation}
S = \frac{1}{2}\ln\left[(2\pi)^n \, \mathrm{det}(\boldsymbol{C}) \, \exp(n)\right],
\end{equation}
for the Shannon-entropy and
\begin{equation}
S_\alpha = \frac{1}{2}\ln\left[(2\pi)^n \, \mathrm{det}(\boldsymbol{C}) \, \alpha^\frac{n}{\alpha-1}\right],
\end{equation}
for the R{\'e}nyi-entropy, such that the univariate case is recovered for $n=1$ and $\mathrm{det}(\boldsymbol{C}) = \sigma^2$. The two definitions are consistent as in the limit $\alpha\rightarrow 1$, the expression $\alpha^\frac{n}{\alpha-1}$ converges to $\exp(n)$. The difference between Shannon- and R{\'e}nyi-entropies for Gaussian distributions with identical covariances is given by an additive term,
\begin{equation}
\Delta_n(\alpha) = S_\alpha - S = \frac{1}{2}\left(\ln\left(\alpha^\frac{n}{\alpha-1}\right) - n\right)
\end{equation}
which is depicted in Fig.~\ref{fig_entropy_scaling}, showing that in particular Bhattacharyya-entropies \citep{Bhattacharyya_original}, where $\alpha=1/2 < 1$, will always be larger than Shannon-entropies. This trend becomes stronger with increasing number of random variables $n$. Typical numbers in cosmology with a $\Lambda$CDM- or $w$CDM-model with 7 or 8 parameters would then be $\Delta_7(1/2) \simeq 1.35$ and $\Delta_8(1/2) \simeq 1.54$.

Remarkably enough, both entropies are measures of the logarithmic volume of the parameter space bounded by the $1\sigma$-contour, implying that the dark energy figure of merit is in fact an inverse information entropy. Interestingly, the entropies are defined as $\mathrm{det}(\boldsymbol{C})$ is always strictly positive for the covariance matrix $C = \bra\boldsymbol{x}\otimes\boldsymbol{x}\ket$ as a consequence of Gram's inequality, while $\mathrm{det}(\boldsymbol{x}\otimes \boldsymbol{x})$ without averaging would be exactly zero. By the choice of the natural logarithm, the unit of entropy is nat.

Clearly, the information entropies $S$ and $S_\alpha$ are inversely proportional to $\ln\det(\boldsymbol{F})$ and one should expect similar relations with the measures $\mathrm{tr}(\boldsymbol{F}) = F^\mu_{\ \ \mu}$ and $\mathrm{tr}(\boldsymbol{F}^2) = F_{\mu\nu}F^{\mu\nu}$ too. As explained before, the H{\"o}lder-inequality and the inequality of the geometric and arithmetic mean provide bounds on the information entropy $S$ for both the Shannon- and R{\'e}nyi-definition in terms of trace invariants of the Fisher-matrix. Additivity in the case of statistical independence is a defining property of information entropies that makes them useful for describing the information content. They share this property with Fisher-matrices for the case of statistically independent probes, i.e. $\boldsymbol{F} = \boldsymbol{F}^{(1)} + \boldsymbol{F}^{(2)}$ implies $S_\alpha = S^{(1)}_\alpha + S^{(2)}_\alpha$ as a consequence of $p(\boldsymbol{x}) = p^{(1)}(\boldsymbol{x}) \: p^{(2)}(\boldsymbol{x})$. For cases with statistical non-independence, additivity of the entropies does not hold and therefore one defines relative entropies between two distributions, also referred to as divergences. For the Shannon-entropy, there is the Kullback-Leibler-divergence $\Delta S$ \citep{kullback1951},
\begin{equation}
\Delta S = D_\mathrm{KL} = \int\dd^n x\:p(\boldsymbol{x})\ln\frac{p(\boldsymbol{x})}{q(\boldsymbol{x})},
\end{equation}
where \citet{baez_bayesian_2014} provide a link to Bayesian statistics, and a more general class of $\alpha$-divergences $\Delta S_\alpha$ for R{\'e}nyi-entropies \citep{van_erven_renyi_2014},
\begin{equation}
\Delta S_\alpha = \frac{1}{\alpha-1}\ln\int\dd^n x\:p(\boldsymbol{x})\left(\frac{p(\boldsymbol{x})}{q(\boldsymbol{x})}\right)^{\alpha-1}
\end{equation}
between two multivariate distributions $p(\boldsymbol{x})$ and $q(\boldsymbol{x})$.

Likewise, relative entropies would be invariant under transformation of the random variables, whereas absolute entropies would not. In fact, they do depend on the choice of parameterisation and even on the choice of units for the parameters, which in particular is less relevant in cosmology as almost all parameters are defined in a dimensionless way, with $H_0$ or $\chi_H = c/H_0$ being notable exceptions. Indeed, under an invertible reparameterisation with a nonzero Jacobian determinant $\det(\partial y^\nu/\partial x_\mu)$, both the Shannon-entropy $S$ and, surprisingly, the R{\'e}nyi-entropy $S_\alpha$ too acquire the identical additive term $\ln\mathrm{det}(\partial y^\nu/\partial x_\mu)$, if the transformation is affine, $y_\nu = A_{\hphantom{\mu}\nu}^\mu x_\mu + b_\nu$ with a constant $A_{\hphantom{\mu}\nu}^\mu$ and $b_\nu$ corresponding to a change in units and a shift of the mean.

In our application, we would like to compute the entropy difference between the posterior $p(\boldsymbol{x}) = p(\boldsymbol{x}|M,\boldsymbol{D})\propto \mathcal{L}(\boldsymbol{D}|\boldsymbol{x},M)\pi(\boldsymbol{x}|M)$ which includes the information provided by measurement and the prior $q(\boldsymbol{x}) = \pi(\boldsymbol{x}|M)$, which reflects the state of knowledge before the data has been taken: Those distribution could either be of specific shape if a previous measurement has already constrained the parameters in question, originate from theory or be chosen in a non-committal way \citep{handley_maximum_2018}. We would like to point out that the entropy divergences $\Delta S$ and $\Delta S_\alpha$ are not symmetric in interchanging prior and posterior, and that the definition of relative entropy does not admit transitivity when combining multiple independent data sets $\boldsymbol{D}_1,\ldots,\boldsymbol{D}_n$, i.e. in cases where $p(\boldsymbol{x}|M,\boldsymbol{D}) = \mathcal{L}(\boldsymbol{D}_1|\boldsymbol{x},M)\cdots\mathcal{L}(\boldsymbol{D}_n|\boldsymbol{x},M)\pi(\boldsymbol{x}|M)$. In fact, if $\Delta S(1)$ is the entropy divergence between the posterior $\mathcal{L}(\boldsymbol{D}_1|\boldsymbol{x},M)\pi(\boldsymbol{x}|M) = \mathcal{L}_1\pi$ and the prior $\pi(\boldsymbol{x}|M)$,
\begin{equation}
\Delta S(1) =
\int\dd^n x\: \mathcal{L}_1\pi\ln\frac{\mathcal{L}_1\pi}{\pi} =
\int\dd^n x\:\mathcal{L}_1\pi\ln\mathcal{L}_1,
\end{equation}
where we suppress the dependence on the parameters, $\boldsymbol{x}$, for notational compactness. Now, $S(2)$ is the corresponding difference between $\mathcal{L}_1\mathcal{L}_2\pi$ and $\pi$,
\begin{equation}
\Delta S(2) =
\int\dd^n x\:\mathcal{L}_1\mathcal{L}_2\pi\ln\frac{\mathcal{L}_1\mathcal{L}_2\pi}{\pi} =
\int\dd^n x\:\mathcal{L}_1\mathcal{L}_2\pi\ln\mathcal{L}_1\mathcal{L}_2
\end{equation}
one can define the entropy decrease $\Delta S(12)$ gained by including the data set $\boldsymbol{D}_2$ and adding the likelihood $\mathcal{L}_2$ to the state of knowledge $\mathcal{L}_1\pi$ obtained from the data set $\boldsymbol{D}_1$,
\begin{equation}
\Delta S(12) =
\int\dd^n x\:\mathcal{L}_1\mathcal{L}_2\pi\ln\frac{\mathcal{L}_1\mathcal{L}_2\pi}{\mathcal{L}_1\pi} =
\int\dd^n x\:\mathcal{L}_1\mathcal{L}_2\pi\ln\mathcal{L}_2.
\end{equation}
With these definitions, one sees that $\Delta S(2) \neq \Delta S(1) + \Delta S(12)$. Because of this and due to statistical non-independence of cosmological probes, we compute all entropies from the effective Fisher-matrix combining all probes into a single Gaussian likelihood. The same issue appears in the case of R{\'e}nyi-entropies $\Delta S_\alpha$ in an identical way.

Using the inverse identification, i.e. setting $p(\boldsymbol{x}) = \pi(\boldsymbol{x}|M)$ and $q(\boldsymbol{x}) \propto \mathcal{L}(\boldsymbol{D}|\boldsymbol{x},M)\pi(\boldsymbol{x}|M)$, i.e. computing the entropy divergence of the prior relative to the posterior yields an interesting result. Now, the entropy divergence quantifies by how much the entropy will decrease by acquiring new data, i.e. by how much the entropy of the posterior will be different relative to that of the prior. At first sight, one might think that  entropies are then additive for statistically independent likelihoods, $\mathcal{L} = \prod_i\mathcal{L}_i$, 
\begin{equation}
\Delta S = 
\int\dd^n x\:\pi\ln\frac{\pi}{\mathcal{L}\pi} = 
-\int\dd^n x\:\pi\ln\mathcal{L} = 
-\sum_i\int\dd^n x\:\pi\ln\mathcal{L}_i,
\end{equation}
but the evidence-term $\int\dd^n x\:\mathcal{L}(x_\mu)\pi(x_\mu)$ needed for a correctly normalised posterior in fact breaks additivity, as
\begin{equation}
\Delta S(1) = -\int\dd^nx\:\pi\ln\mathcal{L}_1 + \ln\int\dd^nx\:\mathcal{L}_1\pi
\end{equation}
with the renormalised posterior,
\begin{equation}
q(\boldsymbol{x}) = \frac{\mathcal{L}(\boldsymbol{x})\pi(\boldsymbol{x})}{\int\dd^nx\:\mathcal{L}(\boldsymbol{x})\pi(\boldsymbol{x})}
\end{equation}
is not contained in the expression
\begin{equation}
\Delta S(2) = -\int\dd^nx\:\pi\ln\mathcal{L}_1 - \int\dd^nx\:\pi\ln\mathcal{L}_2 + \ln\int\dd^nx\:\mathcal{L}_1\mathcal{L}_2\pi.
\end{equation}

If there are no tensions between the likelihoods and if they are of Gaussian shape, one can find analytic relations for the relative Shannon-entropy $\Delta S$,
\begin{equation}
\Delta S = \frac{1}{2}\left[\ln\frac{\mathrm{det}(\boldsymbol{F})}{\mathrm{det}(\boldsymbol{G})} - n + F^{-1}_{\mu\nu}G^{\mu\nu}\right]
\label{eqn_kl_divergence_nobias}
\end{equation}
now expressed in terms of the Fisher-matrices $F_{\mu\nu}$ and $G_{\mu\nu}$ of the posterior and the prior, respectively, as well as for the relative R{\'e}nyi-entropy $\Delta S_\alpha$,
\begin{equation}
\Delta S_\alpha = \frac{1}{2}\frac{1}{\alpha-1}\ln\left[\frac{\mathrm{det}^{\alpha}(\boldsymbol{F})}{\mathrm{det}^{\alpha-1}(\boldsymbol{G})\:\mathrm{det}(\boldsymbol{A})}\right].
\end{equation}
Both relationships yield $\Delta S = \Delta S_\alpha = 0$ if $\boldsymbol{F}=\boldsymbol{G}$. It is quite illustrative to substitute $\ln\mathrm{det}(\boldsymbol{F}) = \ln\mathrm{tr}(\boldsymbol{F})$, yielding
\begin{equation}
\Delta S = \frac{1}{2} \, \Bigg[\left(\ln(F)^\mu_\mu - F^{-1}_{\mu\nu}F^{\mu\nu}\right) - \left(\ln(G)^\mu_\mu - F^{-1}_{\mu\nu}G^{\mu\nu}\right) \Bigg]
\end{equation}
such that $\Delta S$ becomes $\bra\Delta\chi^2\ket/2$ for Gaussian likelihoods as $\mathcal{L}\propto\exp(-\chi^2/2)$, where we substituted $F_{\mu\nu}^{-1}F^{\mu\nu} = n$ for symmetry. The analogous relation for the relative R{\'e}nyi-entropy $\Delta S_\alpha$ is
\begin{equation}
\Delta S_\alpha = \frac{1}{2}\frac{1}{\alpha-1}\left[\alpha\ln(F)^\mu_\mu + (1-\alpha)\ln(G)^\mu_\mu - \ln(A)^\mu_\mu\right]
\label{eqn_relative_renyi}
\end{equation}
with
\begin{equation}
A_{\mu\nu} = \alpha F_{\mu\nu} +(1-\alpha) G_{\mu\nu},
\end{equation}
where one recovers the convexity condition for the matrix-valued logarithm. Again, application of de l'H{\^o}pital's rule for evaluating the limit $\alpha\rightarrow 1$ recovers $\Delta S$ from $\Delta S_\alpha$. It is not straightforward to find general interpretations of eqn.~(\ref{eqn_relative_renyi}) for arbitrary $\alpha$. In the Shannon case, one finds for $S$ the ratio between the logarithmic volumes of the two likelihoods and the asymmetry of the relative entropy is ensured by the fact that $F^{-1}_{\mu\nu}G^{\mu\nu} \neq F^{\mu\nu}G^{-1}_{\mu\nu}$. One would find a symmetric expression for the R{\'e}nyi-entropy $\Delta S_\alpha$ if $\alpha=1/2$, the Bhattacharyya-entropy, in accordance with the definition in this particular case,
\begin{equation}
\Delta S_\alpha = 2\ln\int\dd^n x\:\sqrt{p(\boldsymbol{x})q(\boldsymbol{x})},
\end{equation}
becoming symmetric, with equal prefactors for $F_{\mu\nu}$, $G_{\mu\nu}$ and $A_{\mu\nu}=(F_{\mu\nu}+G_{\mu\nu})/2$. We will come back to this in Sect.~\ref{sect_evidence}, when discussing the relationship between Bayes-evidence and information entropy.

\section{Large scale structure probes}\label{sect_probes}
As discussed in the previous section, we will assume the data to be given as a collection of spherical harmonic modes. Under the assumption of Gaussian fields, their power spectra entirely determine the statistical properties. In this section, we will briefly describe the probes considered  here and how the corresponding spectra are evaluated. For more details, we refer to \citet{2019RobertCode}, which demonstrates the construction of Fisher-matrices $F_{\mu\nu}$ from the cosmological probes including all non-vanishing cross-correlations that would arise \citep{kitching_3d_2014, nicola_integrated_2016, merkel_parameter_2017}. Here, we emphasis that cross-correlations have a dual influence on the inference process by making the data statistically dependent which would decrease the constraining power. On the other hand, they introduce unique handles on investigating structure formation, for instance, through the sensitivity of the integrate Sachs-Wolfe effect the CMB-LSS-correlations to dark energy. We approximate the covariance through a Gaussian with additional power on small scales due to the  modelling of nonlinear structure formation \citep{hilbert_cosmic_2011, kayo_information_2013, krause_cosmolike_2016}. Also, due to the assumption of a true fiducial model, we do not need to worry about covariance matrix variations \citep{tao_random_2012, paz_improving_2015, reischke_variations_2016}. By the Gaussian assumption, there are no complications arising in relation to covariance matrix estimation \citep{taylor_estimating_2014, sellentin_parameter_2016, sellentin_quantifying_2016, sellentin_insufficiency_2017}.

\subsection{Cosmic Microwave Background}
The spectra of cleaned, full-sky CMB maps are given by a
\begin{equation}
\langle a^{P*}_{\ell m} a^{P'}_{\ell' m'}\rangle \equiv \hat{C}^{PP'}(\ell)=
\left(C^{PP'}(\ell)+ N^P(\ell) \right)\delta_{\ell\ell'}\delta_{mm'}\;,
\end{equation}
where $P=T,E,B$ stands for temperature or the two polarization modes respectively, while $C^{TB}(\ell) = C^{EB}(\ell) = 0$. The noise covariance is given by \citep{knox_determination_1995}
\begin{equation}
N^{P}(\ell)\equiv \langle n^{P*}_{\ell m}n^{P'}_{\ell m}\rangle =
\theta^2_\mathrm{beam}\sigma^2_P\exp\left(\ell(\ell+1)\frac{\theta_\mathrm{beam}^2}{8\mathrm{ln}2}\right)\delta_{PP'}\,.
\end{equation}
with root mean square $\sigma^2_P$ and a Gaussian beam with width $\theta_\mathrm{beam}$.
Stage IV CMB experiments {\citep[e.g.][]{thornton_atacama_2016}} will have a very small instrumental noise allowing for measurements up to $\ell \sim 5000$, especially for the polarisation maps. The spectra of the different components are calculated using the \texttt{hi-CLASS} code \citep{hiCLASS}.

\subsection{Large scale structure}
The modes of any large scale structure probe can be calculated, to first order, as a weighted line-of-sight integral of the modes of the density field
\begin{equation}
A_{\ell m} = \int \mathrm{d}\chi\; W_A(\chi)\delta_{\ell m}(\chi)\;,
\end{equation}
where $\chi$ is the comoving distance and a suitable weighting function $W_A(\chi)$. Corresponding spectra involve integration over Bessel functions due to the spherical basis. However, in the flat sky and Limber approximation, the calculation is simplified greatly and any angular power spectrum is given by \citep{1954ApJ...119..655L}
\begin{equation}
C_{AB}(\ell) = \int\frac{\mathrm{d}\chi}{\chi^2}W_A(\chi)W_B(\chi) P_{\delta\delta} \left(\frac{\ell + 0.5}{\chi},\chi\right)\;.
\end{equation}
Note that the comoving wave vector of a mode $k$ is related to the multipole $\ell$ via $k = (\ell + 0.5)/\chi$ in the Limber projection. We will continue by listing the weight functions of all probes used:

\begin{enumerate}

\item Cosmic shear \citep[for reviews, we refer to][]{bartelmann_weak_2001, hoekstra_weak_2008}:
\begin{equation}
W(\chi) = \frac{3\Omega_\mathrm{m}\chi_H^2}{2a\chi}\int _{\mathrm{min}(\chi,\chi_i)}^{\chi_{i+1}}\mathrm{d}\chi'p(\chi')\frac{\mathrm{d}z}{\mathrm{d}\chi'}\left(1-\frac{\chi}{\chi'}\right)\;,
\end{equation}
with the Hubble radius $\chi_H=c/H$, $i$ the tomographic bin index and the Jacobi determinant $\mathrm{d}z/\mathrm{d}\chi' = H(\chi')/c$ due to the transformation of the redshift distribution $p(z)\mathrm{d}z$ of background galaxies in redshift $z$, which is given by \citep{EuclidStudyReport}
\begin{equation}
p(z) \,\mathrm{d}z \propto z^2\exp\left[-\left(\frac{z}{z_0}\right)^\beta\right]\;.
\end{equation}
Typical parameters for stage IV experiments are $z_0\approx 1$ and $\beta = 3/2$.

\item Galaxy clustering \citep[e.g.][]{baumgart_fourier_1991, feldman_power-spectrum_1994,heavens_spherical_1995}
\begin{equation}
W\left(k,\chi\right) =
    \frac{H(\chi)}{c}\,b(k,\chi)\,p(\chi) \  \mathrm{if} \ \chi\in [\chi_i,\chi_{i+1})\;,
\end{equation}
where $b$ is the galaxy bias \citep[as summarised in][]{desjacques_large-scale_2018} for which we assume \citep{ferraro_wise_2015}:
\begin{equation}
b(\chi) = b_0\,[1+z(\chi)]\;,
\end{equation}
with a free positive parameter $b_0$.

\item Lensing of the CMB \citep[e.g.][]{hirata_reconstruction_2003,lewis_weak_2006}:
\begin{equation}
W(\chi) = \frac{\chi_* - \chi}{\chi_*\chi}\frac{H(\chi)}{ca}\;,
\end{equation}
with the comoving distance to the last scattering surface $\chi_*$.

\item Integrated Sachs-Wolfe effect {\citep[iSW,][]{sachs_perturbations_1967}}:
\begin{equation}
W(k,a) = \frac{3}{2\chi_H^3} a^2 E(a) \, F'(k,a)\;,
\end{equation}
where the prime denotes a derivative with respect to $a$ and
\begin{equation}
F(k,a) = 2\frac{D_+(k,a)}{a}\;,
\end{equation}
which is measured in cross-correlation with galaxy clustering and weak lensing.
\end{enumerate}

The noise covariance of cosmic shear and galaxy clustering is given by
\begin{equation}
N_\mathrm{LSS} (\ell) = \sigma^2\frac{n_\mathrm{tomo}}{\bar{n}_\mathrm{gal}}\;\delta_{\ell\ell^\prime},
\end{equation}
with $\sigma = 0.3$ and $\sigma = 1$ for lensing and galaxy clustering, respectively, describing the intrinsic ellipticity of galaxies and the Poissonian fluctuation of galaxy numbers in each bin, $\bar{n}_\mathrm{gal}/n_\mathrm{tomo}$. It should be noted that the tomographic bins are chosen such that the same amount of galaxies, i.e. $\bar{n}_\mathrm{gal}/n_\mathrm{tomo}$, lie in each bin. For CMB-lensing, we assume the noise to be given by the quadratic estimator described in \citet{hu_mass_2002,okamoto_cosmic_2003} using all five non-vanishing estimators involving $T$, $E$ and $B$.

In our analysis, we combine the currently most powerful cosmological probes into a joint likelihood function. Specifically, we start out with spectra of the temperature and polarisation anisotropies in the CMB (labeled as CMB primary), and successively add CMB-lensing, tomographic galaxy clustering (GC) and tomographic weak gravitational shear (WL), while taking account of all possible cross-correlations. As the reference cosmology, we use the \citet{Planck2018}-result. The Fisher-matrices used in this work were computed with the code of \cite{2019RobertIA} where all cross-correlations are taken into account. Currently, to apply this approach to available data would not be fully correct since it is not available the correlation between probes that have overlap areas of scan as it is the case for some surveys of galaxy clustering and weak lensing. Therefore we use these specific Fisher-matrices as a proof of concept, that is, to assess relations between the Bayesian statistics methods and information theory measures.

\section{entropy decrease in probe combination}\label{sect_decrease}
In this section we would like to see how the statistical uncertainty in a $\Lambda$CDM- or $w$CDM-cosmology is reduced by combining cosmological probes. Starting from constraints from the temperature and polarisation anisotropies of the cosmic microwave background, we add successively gravitational lensing of the CMB, galaxy clustering and weak gravitational lensing, i.e. the large-scale structure probes ordered by decreasing redshift. In doing that, we are considering all nonzero cross-correlations in the data covariance, most notably the integrated Sachs-Wolfe effect between the CMB-temperature, any low-redshift tracer of the large-scale structure, and the nonzero cross-correlations between the galaxy density and weak lensing.

At some point, the distribution will become very narrow such that their entropies, irrespective of the Shannon- or R{\'e}nyi-definition, will become negative. One can explain this straightforwardly by considering a one-dimensional Gaussian distribution with variance $\sigma^2$, where the relevant term in both entropy definitions is $\ln(\sigma^2)$ which tends towards $-\infty$ as $\sigma^2\rightarrow 0$. The $\delta_D$-distribution with perfect knowledge of the parameters has infinite negative entropy (if one considers $\delta_D(x)$ as a distribution in the probabilistic sense), and not zero as a consequence of the continuum limit.

First we quantify the absolute entropies $S$ and $S_\alpha$ for the four cosmological data sets separately and put them into relation with other measures of total error that can be directly derived from the Fisher-matrix such as $\mathrm{tr}(\boldsymbol{F})$ and $\mathrm{tr}(\boldsymbol{F}^2)$, where the inequality
\begin{equation}
\ln\mathrm{det}(\boldsymbol{F}) = \mathrm{tr}\ln \boldsymbol{F} \geq \ln\mathrm{tr}(\boldsymbol{F})
\end{equation}
is obeyed as should be expected from a positive definite and symmetric Fisher-matrix $F_{\mu\nu}$. As such, the entropies are in fact not only scaling with Fisher-invariants but are bounded by them as well, keeping in mind that $2S = n(\ln(2\pi)+1) - \ln\mathrm{det}(\boldsymbol{F})$.

Specifically, absolute Shannon-entropies $S$ and R{\'e}nyi-entropies $S_\alpha$ for $\alpha=1/2$ are listed in Table~\ref{table_absolute_lambdacdm} for a $\Lambda$CDM cosmology and in Table~\ref{table_absolute_wcdm} for a $w$CDM-cosmology. Clearly, for both cosmological models, the cosmic microwave background is the primary source of information, followed by galaxy clustering and weak lensing, and with CMB-lensing adding the smallest amount of information. Comparing the two cosmological models, the entropies in $\Lambda$CDM are smaller, reflecting the reduced parameter space in comparison to $w$CDM, leading to tighter constraints, smaller entries in the parameter covariance matrix and in consequence, of the information entropies. The Shannon- and R{\'e}nyi-entropies are related for the Gaussian distributions by a fixed factor, for which we have chosen to compute the case for $\alpha = 1/2$.

\begin{table}
\begin{tabular}{lrr}
\hline\hline
probe & Shannon-entropy $S$ & Bhattacharyya-entropy $S_\alpha$ \\
\hline
CMB 				& -28.50 & -27.53 \\
CMB-lensing			&  -8.58 &  -7.57 \\
galaxy clustering 	& -18.90 & -18.02 \\
weak lensing 		& -16.51 & -15.55 \\
\hline
\end{tabular}
\caption{Absolute Shannon- and Bhattacharyya-entropies $S$ and $S_\alpha$, $\alpha=1/2$, in units of nats, for the likelihood of a $\Lambda$CDM-model, computed from the Fisher-matrices.}
\label{table_absolute_lambdacdm}
\end{table}

\begin{table}
\begin{tabular}{lrr}
\hline\hline
probe & Shannon-entropy $S$ & Bhattacharyya-entropy $S_\alpha$ \\
\hline
CMB 				& -31.04 & -30.08 \\
CMB-lensing			&  -8.39 &  -7.43 \\
galaxy clustering 	& -22.19 & -21.22 \\
weak lensing 		& -18.51 & -17.54 \\
\hline
\end{tabular}
\caption{Absolute Shannon- and Bhattacharyya-entropies $S$ and $S_\alpha$, $\alpha=1/2$, in units of nats, characterising the likelihood of a $w$CDM-model where $w$ is a constant allowed to be different from -1.}
\label{table_absolute_wcdm}
\end{table}

\begin{figure}
\centering
\includegraphics[scale=0.45]{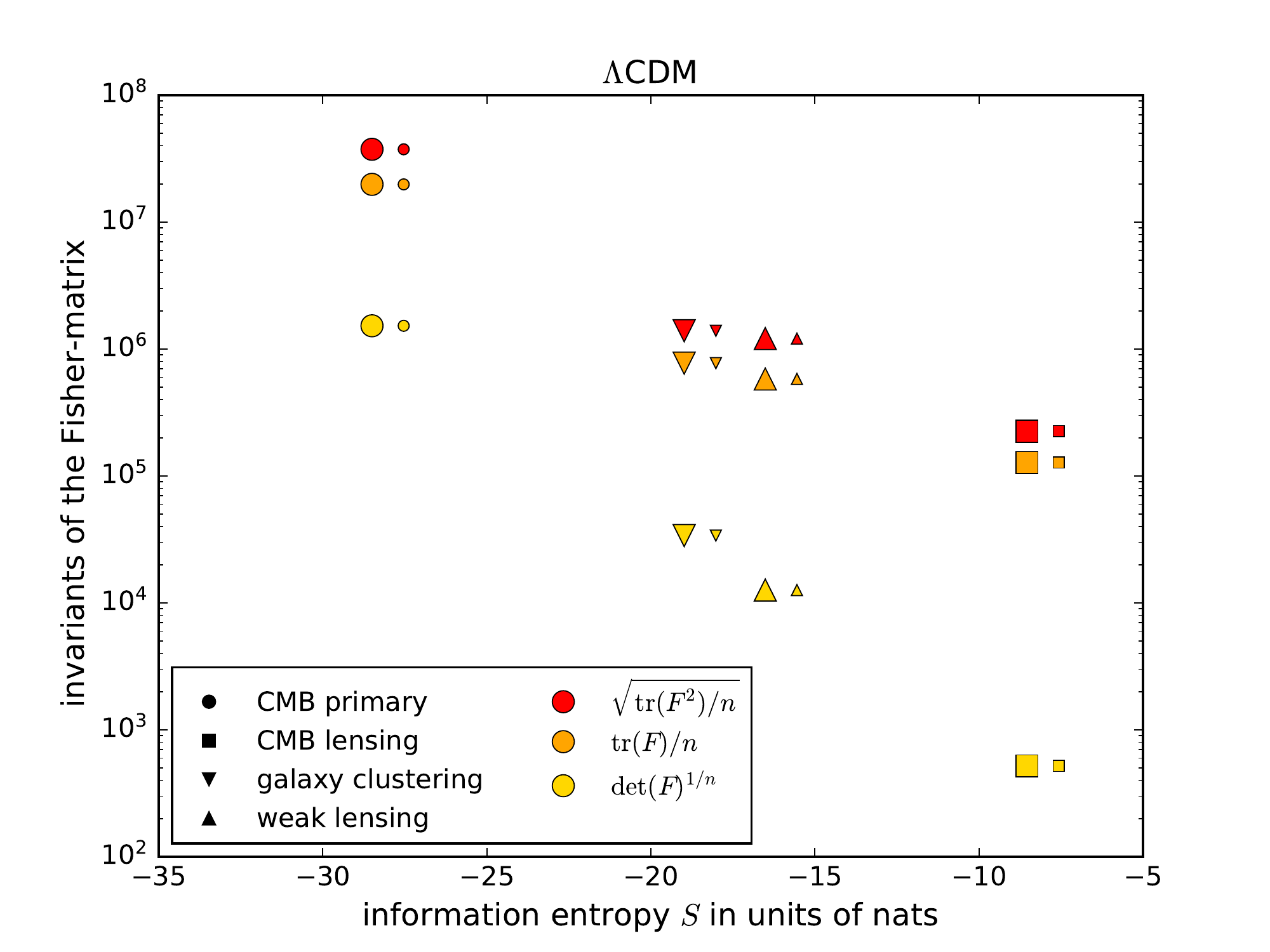}
\caption{Absolute Shannon-entropy $S$ (large symbols) and Bhattacharyya-entropye $S_\alpha$, $\alpha=1/2$ (small symbols) in units of nats for the likelihood of a $\Lambda$CDM-cosmology, constrained through primary CMB-fluctuations, CMB-lensing, galaxy clustering and weak lensing individually, plotted against $\mathrm{tr}(\boldsymbol{F})/n$, $(\mathrm{tr}(\boldsymbol{F}^p)/n)^{1/p}$, $p=2$, and $\mathrm{det}(\boldsymbol{F})^{1/n}$, with $n=5$.}
\label{fig_absolute_entropy_lcdm}
\end{figure}

\begin{figure}
\centering
\includegraphics[scale=0.45]{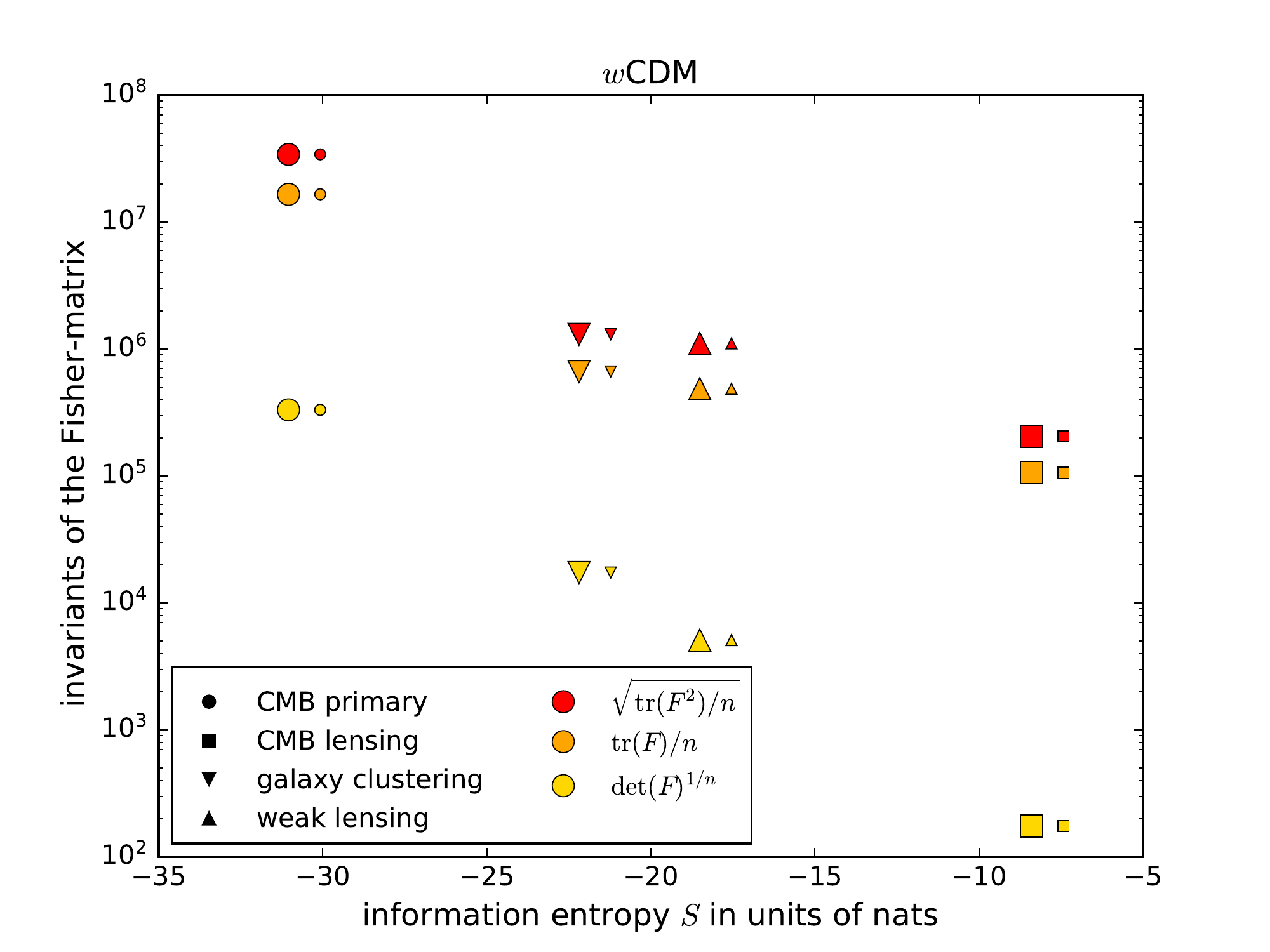}
\caption{Absolute Shannon-entropy $S$ (large symbols) and Bhattacharyya-entropy $S_\alpha$, $\alpha=1/2$ (small symbols) in units of nats for the likelihood of a $w$CDM-cosmology, constrained through primary CMB-fluctuations, CMB-lensing, galaxy clustering and weak lensing individually, plotted against $\mathrm{tr}(\boldsymbol{F})/n$, $(\mathrm{tr}(\boldsymbol{F}^p)/n)^{1/p}$, $p=2$, and $\mathrm{det}(\boldsymbol{F})^{1/n}$, with $n=6$.}
\label{fig_absolute_entropy_wcdm}
\end{figure}

In contrast to absolute entropies, relative entropies $\Delta S$ and $\Delta S_\alpha$ are independent under transformations of the random variable, so in particular the choice of units does not matter. In cosmology, however, there is the particular situation that most of the cosmological parameters are defined in a dimensionless way, such that it is sensible to compare absolute entropies directly. We give in Fig.~\ref{fig_absolute_entropy_lcdm} the total entropy of all cosmological probes individually, and show their scaling with $\mathrm{tr}(\boldsymbol{F})/n$, $(\mathrm{tr}(\boldsymbol{F})/n)^{1/p}$ and $\ln\mathrm{det}(\boldsymbol{F})/n$. Not surprisingly, information entropies show in fact that they scale with trace-invariants of the Fisher-matrix. Clearly, the primary CMB has the highest information content for a $\Lambda$CDM-model, followed by galaxy clustering, weak lensing and CMB-lensing, in that particular order. In addition, the inequalities~\ref{eqn_hoelder} and~\ref{eqn_arith_geo} are clearly fulfilled. Similary, Fig.~\ref{fig_absolute_entropy_wcdm} shows for the same cosmological probes their respective information content for a $w$CDM cosmology. Strong degeneracies between the parameters can give rise to small values for the determinant and therefore large values for $S$.

\section{information entropies and measures of statistical error}\label{sect_statistics}
Information entropies and invariants of the Fisher-matrix are both measures of the total statistical uncertainty, and as such one should expect a relation between $S$ (or generally $S_\alpha$) with $\mathrm{tr}(\boldsymbol{F})$, $\mathrm{tr}(\boldsymbol{F}^2)$ and $\mathrm{det}(\boldsymbol{F})$ at every stage of combining cosmological probes.

While it is clear that in the case of an uncorrelated multivariate Gaussian distribution with covariance $C_{\mu\nu}\propto\delta_{\mu\nu}$,
the entropies of the individual distributions add, $S = \sum_\mu S_\mu$, the same does not hold if correlations are present. In fact, the total entropy is bounded by the conditional and marginal variances, respectively. That is, for both Shannon- and R{\'e}nyi-entropies the conditional error results from the corresponding inverse entry of the Fisher-matrix, $\sigma^2_{\mu,c} = (F_{\mu\mu})^{-1}$, such that with $S^{(c)}_\mu\propto\sigma_\mu^2$, one obtains (ignoring non-relevant prefactors) $\exp(-S^{(c)}_\mu) = F_{\mu\mu}$. Using the Hadamard-inequality one then finds $\exp(-S) = \mathrm{det}(F)\leq\prod_\mu F_{\mu\mu} = \prod_\mu \sigma^2_{\mu,c} = \prod_\mu\exp(-S^{(c)}_\mu) = \exp\left(\sum_\mu S^{(c)}_\mu\right)$, such that $S\geq \sum_\mu S^{(c)}_\mu$ for conditional variances. Conversely, the marginalised variance is computed from the inverse Fisher-matrix, $\sigma^2_{\mu,c} = (\boldsymbol{F}^{-1})_{\mu\mu}$, as well as $\mathrm{det}(\boldsymbol{F}^{-1}) = 1/\mathrm{det}(\boldsymbol{F})$ implying $\exp(S) = \mathrm{det}(\boldsymbol{F}^{-1})$. Then, $\exp(S) = \mathrm{det}(\boldsymbol{F}^{-1}) \leq \prod_\mu (\boldsymbol{F}^{-1})_{\mu\mu} = \prod_\mu \sigma^2_{\mu,m} = \prod_\mu \exp(-S^{(m)}_\mu) = \exp\left(\sum_\mu S^{(m)}_\mu\right)$, and from that $S \leq \sum_\mu S^{(m)}_\mu$, i.e. in summary $\sum_\mu S_\mu^c\leq S \leq \sum_\mu S^{(m)}_\mu$, where equality is given for the uncorrelated case.

Additionally, the Cram{\'e}r-Rao-inequality asserts that the estimated variance of a distribution is bounded by the Fisher-matrix from below, where equality between the variances $\sigma^2$ and $\boldsymbol{F}^{-1}$ is only given for a Gaussian distribution. If one were to estimate the covariance matrix from a non-Gaussian distribution, the resulting variance would be larger than that of a Gaussian distribution for the same Fisher-matrix, and assigning an entropy to that covariance through $S\propto\ln\mathrm{det}(\boldsymbol{C})$ would yield a larger result than $-\ln\mathrm{det}(\boldsymbol{F})$. This statement is not in contradiction with the property of the Gaussian distribution to maximise $S$ for a given covariance, because the actual value of $S$ depends on the shape of the distribution and has to be either computed from the functional shape or be estimated from data, through $S = -\int\dd^nx\:p\ln p$ or $S_\alpha = -\int\dd^nx\:pp^{\alpha-1}/ (\alpha-1)$.

It is a standard derivation to show by functional extremisation $\delta S=0$ of $S = -\int\dd x\: p(x)\ln p(x)$ while incorporating the boundary conditions $\int\dd x\:p(x) = 1$ and $\int\dd x\: p(x) x^2 = \sigma^2$ with Lagrange-multipliers that the Gaussian distribution is in fact the one with the largest possible entropy for fixed variance. We would like to point out that the Gaussian distribution is likewise the solution if one fixes the Fisher-matrix $F = \bra(\partial\ln p)^2\ket = \int\dd x\: p(\partial\ln p)^2$. Formulating the entropy functional as the averaged logarithmic curvature,
\begin{equation}
S = -\int\dd x\:p\ln p + \lambda\left[\int\dd x\:p - 1\right] + \mu\left[\int\dd x\:p (\partial\ln p)^2 - F\right]
\end{equation}
yields as a solution to $\delta S = 0$ the differential equation
\begin{equation}
\ln p(x) + 1 + \lambda + \mu\frac{\partial^2p}{p} = 0
\end{equation}
using $\partial^2p/p = -(\partial\ln p)^2$, which is solved by the Gaussian distribution $p(x) = \exp(-x^2/(2\sigma^2))/\sqrt{2\pi\sigma^2}$ while identifying $F$ with $\sigma^{-2}$.

\begin{figure}
\includegraphics[scale=0.45]{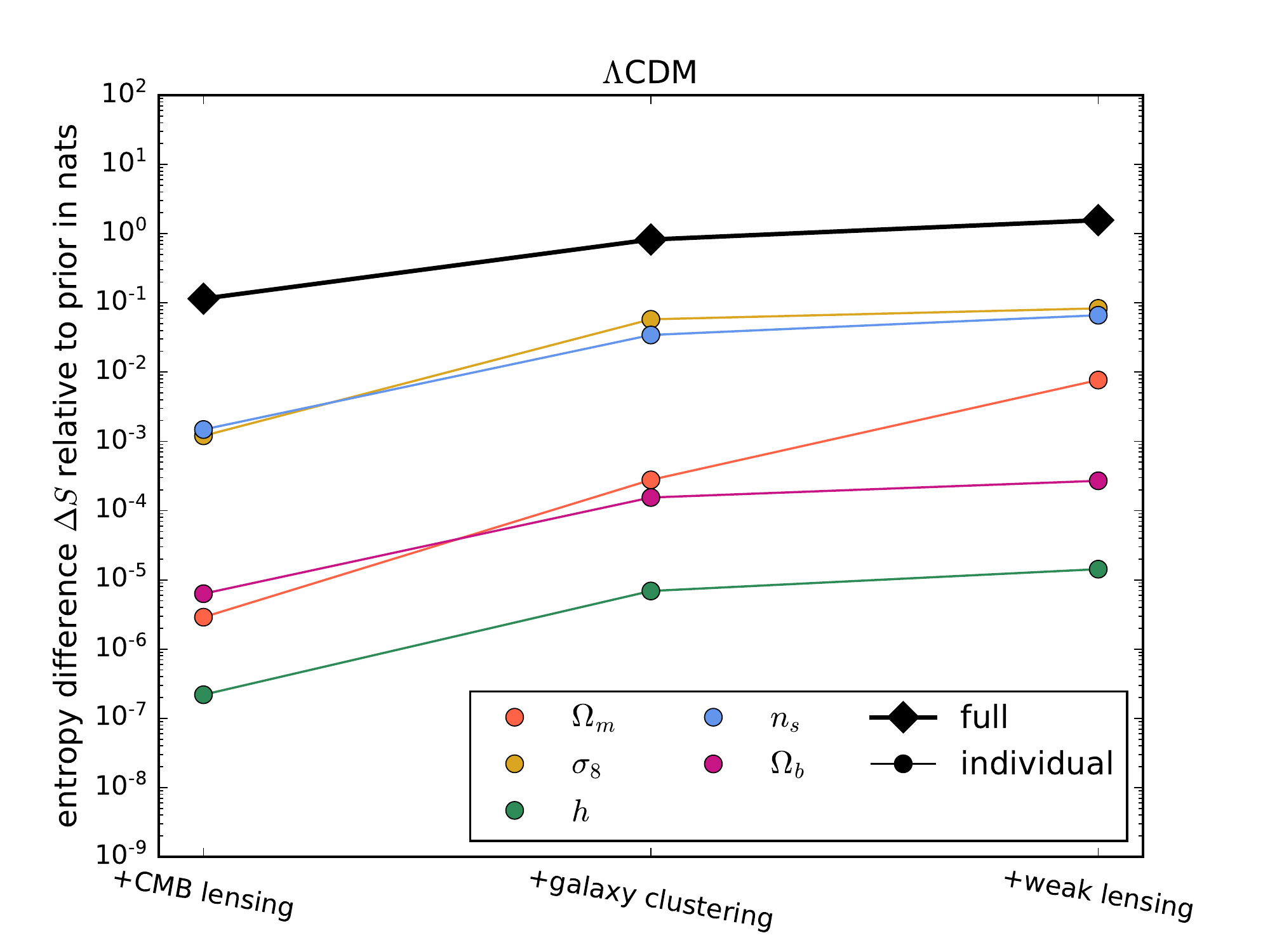}
\caption{Relative entropies $\Delta S$ for the full likelihood of a $\Lambda$CDM-cosmology, and for all $n=5$ parameters individually, both marginalised and conditionalised, for a successive and cumulative probe combination adding CMB-lensing, galaxy clustering and weak lensing to the primary CMB.}
\label{fig_relS-LCDM}
\end{figure}

\begin{figure}
\includegraphics[scale=0.45]{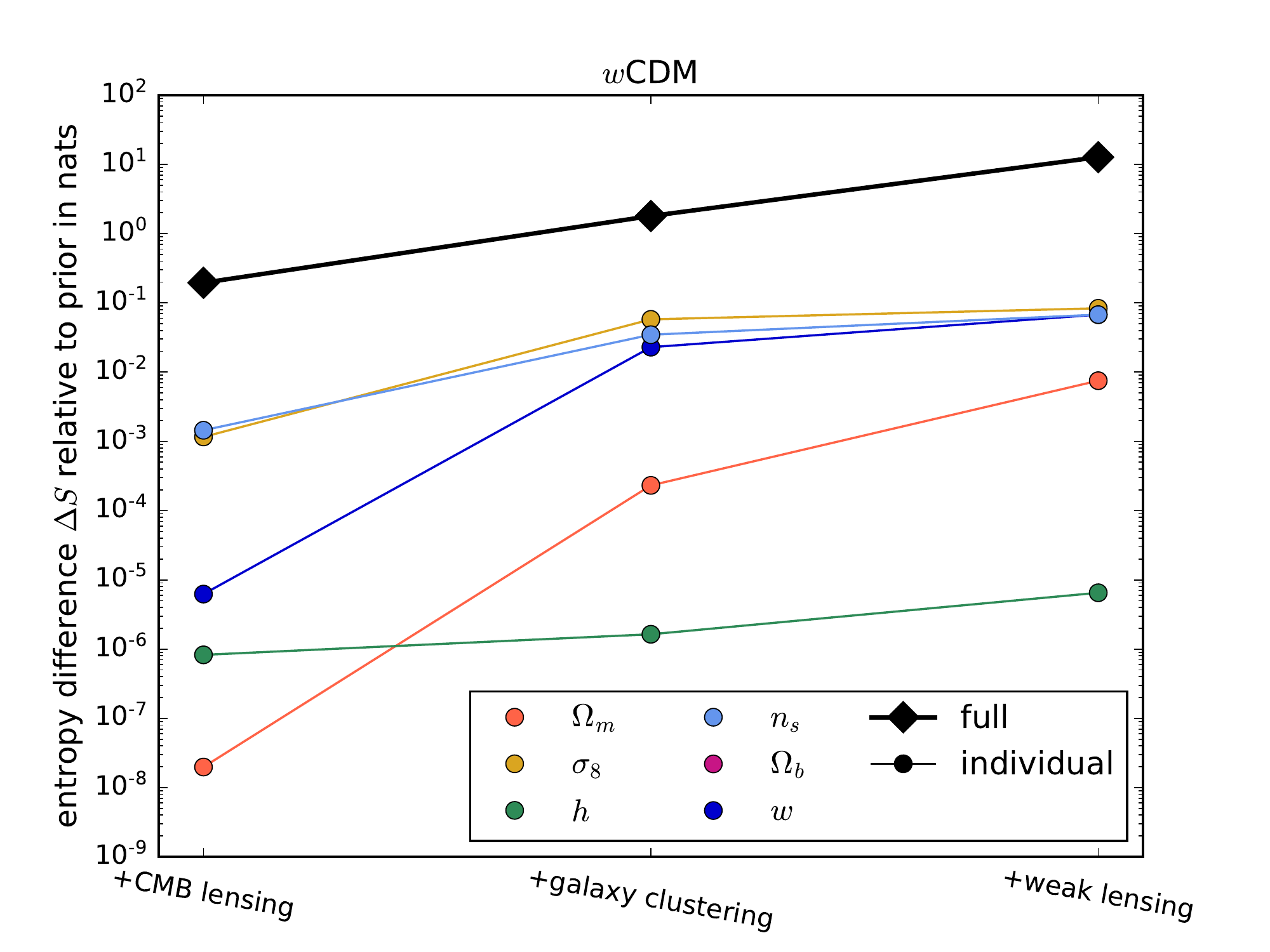}
\caption{Relative entropies $\Delta S$ for the full likelihood of a $w$CDM-model, and for the individual liklihoods for all $n=6$ parameters, both marginalised and conditionalised, for a successive combination of cosmological probes adding CMB-lensing, galaxy clustering and weak lensing to the primary CMB.}
\label{fig_relS-w0CDM}
\end{figure}

We compute information entropy differences for combinations of cosmological data sets and add successively, in order of decreasing redshift, CMB-lensing, tomographic galaxy clustering and tomographic weak gravitational lensing to the primary CMB-fluctuations. Specifically, we compute the resulting combined Fisher-matrix including all cross-correlations and use it to quantify the gain of information of the probe combination over the primary CMB. We quantify the Kullback-Leibler-divergence $\Delta S$ for the full likelihood (where any R{\'e}nyi-entropy difference would differ by a numerical factor depending on the dimensionality of the likelihood). The results are shown in Fig.~\ref{fig_relS-LCDM} and~\ref{fig_relS-w0CDM}, for the Kullback-Leibler divergence $\Delta S$. In addition, we repeat the analysis for individual likelihoods of each individual parameter $\Omega_m$, $\sigma_8$, $h$, $n_s$ and $w$, where the above discussed inequalities for the sum of the conditionalised entropies in comparison to the total entropy become relevant. In contrast to the absolute entropies discussed in the previous chapter, the relative entropies are invariant under reparameterisation of the model and do not rely on a preferred parameterisation. Clearly, there is a reduction in uncertainty achieved through the combination of cosmological probes reflected by smaller absolute information entropies. One observes a more dramatic effect in the $w$CDM-model compared to the $\Lambda$CDM-model. The interpretation of the absolute numbers of $\Delta S$ is not straightforward, as they combine a measure of change of admissible volume of parameter space, $\ln\mathrm{det}F - \ln\mathrm{det}G$ with the measure $(F^{-1}_{\mu\nu}G_{\mu\nu})-n$ sensitive to the relative orientation of the eigensystems of $F_{\mu\nu}$ and $G_{\mu\nu}$, i.e. of the changing degeneracies in probe combination.

\section{entropy increase through systematics}\label{sect_systematics}
Up to this point we always worked under the assumption of unbiased measurements, such that the averaged likelihoods in fact peaked at the fiducial cosmology because the data was on average equal to the theoretical prediction. We will relax this assumption by considering shifts in the likelihood functions of different cosmological probes due to systematical errors. In a previous paper we have defined the figure of bias $Q$ from the Fisher-matrix $F_{\mu\nu}$ and the shifts $\delta_\mu$ of the best-fit point through the quadratic form $Q^2 = \sum_{\mu\nu}F_{\mu\nu}\delta_\mu\delta_\nu$ and showed the relationship to the Kullback-Leibler-divergence $\Delta S = Q^2/2$ if the covariance is unaffected by the systematic. While the interpretation of $Q$ as the systematic error in units of the statistical error is straightforward, it is likewise obvious that there is an effect of systematic errors on relative entropies: In fact, the explicit relationship for the Kullback-Leibler-divergence $\Delta S$ between two Gaussian distributions with Fisher-matrices $F_{\mu\nu}$ and $G_{\mu\nu}$ has a term involving $\delta_\mu$,
\begin{equation}
\Delta S = \frac{1}{2}\left(\delta_\mu G_{\mu\nu} \delta_\nu + \ln\frac{\mathrm{det}(\boldsymbol{F})}{\mathrm{det}(\boldsymbol{G})} - n + {F}^{-1}_{\mu\nu}{G}^{\mu\nu}\right),
\label{eq.DKL}
\end{equation}
which reverts to eqn.~(\ref{eqn_kl_divergence_nobias}) in the case of vanishing tension, $\delta_\mu = 0$. The analogous relationship for the relative R{\'e}nyi-entropy $\Delta S_\alpha$ can be derived to be
\begin{equation}
\begin{split}
\Delta S_\alpha & = \frac{1}{2}\frac{1}{\alpha-1}
\ln\left(\frac{\mathrm{det}^\alpha(\boldsymbol{F})}{\mathrm{det}^{\alpha-1}(\boldsymbol{G})\mathrm{det}(\boldsymbol{A})}\times\right.\\
&\quad\quad\quad\quad\quad\quad\quad \left.\exp\left[-\frac{\alpha}{2}\delta^\kappa\left(F_{\kappa\beta} -\alpha F^\mu_{\kappa}A^{-1}_{\mu\nu}F^\nu_\beta\right) \delta^\beta\right]\right),
\end{split}
\label{eqn_relative_renyi_entropy}
\end{equation}
again with
\begin{equation}
A_{\mu\nu} = \alpha F_{\mu\nu} + (1-\alpha) G_{\mu\nu},
\end{equation}
where the previous relation for the R{\'e}nyi-entropy $\Delta S_\alpha$ is recovered for $\delta_\mu = 0$, as the exponential becomes equal to one. Clearly, in this way the entropy difference becomes a measure of consistency \citep{nicola_consistency_2018} as it is sensitive to differences between parameter values derived with different probes, but in addition there is a dependence on the difference between the errors. Additionally, relative entropies provide a much better characterisation of distributions than $p$-values which are necessarily restricted to univariate distributions.

In the following we will consider three well-known examples of tensions between likelihoods, which we approximate by Gaussian distributions: The tension in the value of the Hubble-Lema{\^i}tre parameter $H_0$ between the CMB and Cepheids, the tension in the $(\Omega_m,\sigma_8)$-plane between the CMB and weak lensing, and intrinsic alignments as a contaminant in weak lensing data as a theoretical forecast. Concerning the interpretation of $\Delta S$ and $\Delta S_\alpha$ it is straightforward to show that the Kullback-Leibler-divergence for two identical Gaussian distributions shifted by $\delta$ is given by $(\delta/(\sqrt{2}\sigma))^2$, such that the square root measures the number of standard deviations by which the Gaussian distributions are displaced relative to each other. This immediately suggests the interpretation of the integrated probability to obtain values larger than the actual bias $\delta$ as
\begin{equation}
p = \frac{1}{2}\mathrm{erf}\left(-\frac{\delta}{\sqrt{2}\sigma}\right), 
\end{equation}
i.e. the $p$-value commonly used in descriptive statistics.

\subsection{Hubble-Lema{\^i}tre parameter $H_0$ from Cepheids and the CMB}
The Hubble-Lema{\^i}tre parameter $H_0$ quantifies the current rate of expansion of the Universe, but values from the CMB \citep{Planck2018} and the local value of $H_0$ \citep{2019Riess} are in mutual significant disagreement, for either measurement as a reference. While from the CMB temperature fluctuations one can infer the cosmological parameter values, assuming a $\Lambda$CDM model, with very good statistical precisions, improvements on the distance ladder using near-infrared Cepheids variables in host galaxies with recent type Ia supernovae reduced the uncertainty on $H_0$ to 2$\%$. Typically this measurement is model-independent as it follows directly from the Hubble-Lema{\^i}tre-law and not dependent on a specific cosmological model. The systematic, whose origin is yet unresolved, shows up as a nonzero Kullback-Leibler divergence $\Delta S$. With CMB as a reference (using the best-fit $h_\mathrm{CMB} = 0.68\pm0.005$ for Planck) to be updated by Cepheids (with $h_\mathrm{Cepheid} = 0.7403\pm0.0142$), this gives a value of $\Delta S\simeq 44$~nats, and has a value of $Q\simeq 12$ for the figure of bias. The discrepancy between $\Delta S$ and $Q^2/2$ indicates differences between the variances reported by the two measurements. 

\begin{figure}
\includegraphics[scale=0.45]{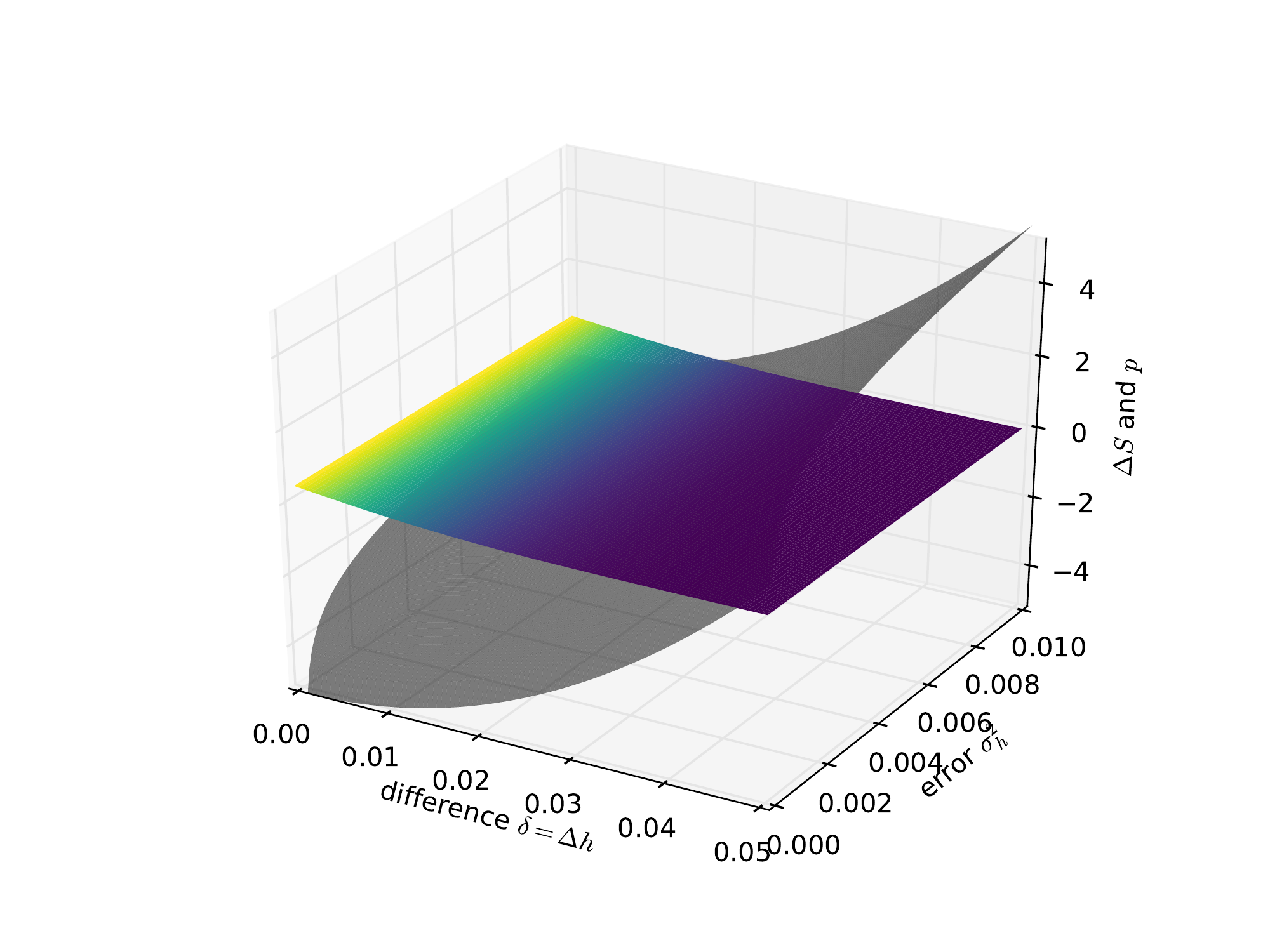}
\caption{Kullback-Leibler divergence $\Delta S$ (gray surface) and $p$-value, green-yellow surface) as a function of difference in $h$ of the Hubble-Lema{\^i}tre $H_0$ parameter between Cepheid- and CMB-measurements and the error $\sigma^2_h$ of the CMB-measurement.}
\label{fig_pvalue}
\end{figure}

It might be interesting to compare the Kullback-Leibler-divergence $\Delta S$ of the two distributions to the $p$-value, which reflects the probability of obtaining values more extreme than the one from the current data set. As the tension in $h$ is predominantly present in Planck's data set, we vary the difference $\delta = h_\mathrm{CMB}-h_\mathrm{Cepheid}$ in the Hubble-Lema{\^i}tre parameter and the error $\sigma^2_h$ of the CMB-measurements. The results on the Kullback-Leibler-divergence $\Delta S$ and the $p$-value is shown in Fig.~\ref{fig_pvalue}. Surely, high values for $\Delta S$ point at large discrepancies between the two distributions as would a low $p$-value, but apart from superficial similarities it is difficult to make general statements.

\subsection{$(\Omega_m,\sigma_8)$-plane from the CMB and weak lensing}
There is as well a long-standing discrepancy between determinations of $\Omega_m$ and $\sigma_8$ from the CMB and from weak gravitational lensing, commonly with lensing preferring smaller values for both parameters relative to the CMB, but with the well-known degeneracy as a lensing essentially determines the product of both parameters. Since the best fit value for $(\Omega_m, \sigma_8)$ does not have a Gaussian uncertainty, we opt to use the parameter 
$S_8 = \sigma_8 \sqrt{\Omega_m /0.3}$ which encapsulates the information of both parameters and its distribution is better approximated by a Gaussian. One obtains an entropy difference $\Delta S\simeq 2.1$~nats for the Kullback-Leibler divergence, between Planck's CMB observation and KiDS's weak lensing data set \cite{KiDS450-2017} with CMB as the reference value, for a fit of a $w$CDM-cosmology to the data with a prior on spatial flatness, indicating a much less severe tension than in the case of $H_0$.

\subsection{$w$CDM and lensing with intrinsic alignments}
Lastly, we quantify the effect of intrinsic alignments in weak lensing data on parameter estimation and the bias that they cause: From a physical point of view, intrinsic alignments are mechanisms related to tidal interaction of galaxies with the large-scale structure, which causes them to have correlated intrinsic shapes, therefore changing fundamental assumption that lensing is the only mechanism to generate shape correlations. When deriving intrinsic ellipticity spectra using tidal shearing for elliptical and tidal torquing for spiral galaxies including the cross correlation that exists between gravitational lensing and the intrinsic shapes of elliptical galaxies, one can derive estimation biases for the $w$CDM parameter set including the dark energy equation of state parameters $w_0$ and $w_a$ \citep{tugendhat_angular_2018} in the parameterisation $w(a) = w_0 + (1-a)w_a$. Expressed in terms of the ratio $\delta/\sigma$, those are in fact significant for a weak lensing survey like Euclid. In this particular case, we work with the approximation that the intrinsic alignments only give rise to a nonzero bias $\delta_\mu$ while keeping the covariance, or equivalently, the Fisher-matrix $F_{\mu\nu}$ fixed, such that the entropy difference recovers the figure of bias $Q^2/2 = F^{\mu\nu}\delta_\mu\delta_\nu$ as all other terms in eqn.~\ref{eq.DKL} vanish: $(F^{-1})_{\mu\nu}G^{\mu\nu}$ becomes $n$, and the logarithm of the ratio of the determinants becomes zero. The numerical value of the Kullback-Leibler-divergence is computed to be $\Delta S\simeq 3\times10^3$~nats if $w$ is constant and $\Delta S\simeq 10^4$~nats if $w$ can evolve linearly with time, for an analysis of Euclid's weak lensing data with 5-bin tomography in the framework of a $w$CDM-cosmology. With the simplified interpretation of $\Delta S$ in this case of constant covariances, $Q$ assumes a value of roughly $30$ in the first and $100$ in the second case, indicating highly significant biases with the interpretation that $Q$ measures the magnitude of the systematic error in units of the statistical error while keeping track of the degeneracy orientation.

In all of these cases, it mattered which likelihood is chosen as the reference for computing the expectation value of $\ln(p(\boldsymbol{x})/q(\boldsymbol{x}))$: One could construct a symmetrised Kullback-Leibler divergence by interchanging the distributions and averaging, but we would like to propose to compute the Wasserstein-metric $\Delta W_2$ \citep{olkin_distance_1982, dowson_frechet_1982},
\begin{equation}
\Delta W_2 = \left|\delta_\mu\right|^2 + \mathrm{tr}\left(\boldsymbol{A}^{-1}+\boldsymbol{B}^{-1}-2(\boldsymbol{BA})^{-1/2}\right),
\end{equation}
for which an analytic expression for Gaussian distributions in terms of the covariances and their Cholesky-decompositions $F_{\mu\nu} = A^\alpha_{\mu}A_{\nu\alpha}$ and $G_{\mu\nu} = B^\alpha_{\mu}B_{\nu\alpha}$ exists. The Wasserstein-metric $\Delta W_2$ is symmetric in both distributions and can serve as an information measure for the dissimilarity between two distributions, as $\Delta W_2 = 0$ for $\delta_\mu = 0$ and $F_{\mu\nu} = G_{\mu\nu}$, and it is remarkable that the tension $\delta_\mu$ enters through its Euclidean norm. There is no obvious relationship between the Wasserstein-metric and Kullback-Leibler- or $\alpha$-divergences, and neither for the Bhattacharyya-entropy, which among all R{\'e}nyi-entropies is the only symmetric one.

\section{Bayesian evidence and information entropy}\label{sect_evidence}
Bayesian evidence as a criterion for model selection provides a trade-off between the size of the statistical errors and the model complexity. It is straightforward to show that for a Gaussian likelihood $\mathcal{L}(\boldsymbol{D}|\boldsymbol{x},M)$ with a Fisher-matrix $F_{\mu\nu}$ and a Gaussian prior $\pi(\boldsymbol{x}|M)$ with the inverse covariance $P_{\mu\nu}$ the evidence $p(\boldsymbol{D}|M)$ is given by
\begin{equation}
p(\boldsymbol{D}|M) = \sqrt{\frac{\mathrm{det}(\boldsymbol{F})\mathrm{det}(\boldsymbol{P})}{(2\pi)^n\mathrm{det}(\boldsymbol{F}+\boldsymbol{P})}},
\end{equation}
which implies a scaling $\propto \pi^{-n}/2$ disfavouring models with high complexity.

The expression for the evidence $p(\boldsymbol{D}|M)$ changes to
\begin{equation}
\begin{split}
p(\boldsymbol{D}|M) &= \sqrt{\frac{\mathrm{det}(\boldsymbol{F})\mathrm{det}(\boldsymbol{P})}{(2\pi)^n\mathrm{det}(\boldsymbol{F}+\boldsymbol{P})}}\times\\
&\quad\quad\quad\quad
\exp\left\{-\frac{1}{2}\delta^\mu\left[P_{\mu\nu} - P_{\mu\alpha}(F+P)^{-1}_{\alpha\beta}P^\beta_{\nu}\right]\delta^\nu\right\},
\end{split}
\end{equation}
if likelihood and prior are displaced by $\delta_\mu$ relative to each other. Comparing this expression with eqn.~(\ref{eqn_relative_renyi_entropy}) shows that the two expressions are related to each other if $\alpha = 1-\alpha$, i.e. if $\alpha = 1/2$, which is know as Bhattacharyya-entropy. For this particular case, the R{\'e}nyi-entropy would weigh both likelihood and prior equally,
\begin{equation}
\Delta S_\alpha = 2\ln\int\dd^n x\:\sqrt{\mathcal{L}(\boldsymbol{D}|\boldsymbol{x},M) \pi(\boldsymbol{x}|M)},
\end{equation}
with the natural bound $\sqrt{\mathcal{L}(\boldsymbol{D}|\boldsymbol{x},M)\pi(\boldsymbol{x}|M)}\leq [\mathcal{L}(\boldsymbol{D}|\boldsymbol{x},M)+\pi(\boldsymbol{x}|M)]/2$ given by the inequality of the geometric and arithmetic mean, such that $\Delta S_\alpha\geq 0$ as both $\mathcal{L}(\boldsymbol{D}|\boldsymbol{x},M)$ and $\pi(\boldsymbol{x}|M)$ are normalised. Because the logarithm is a concave function, one can use Jensen's inequality to write $\ln\int\dd^n x\:\sqrt{\mathcal{L}(\boldsymbol{D}|\boldsymbol{x},M) \pi(\boldsymbol{x}|M)} \geq \int\dd^n x\:\ln\sqrt{\mathcal{L}\pi} = \int\dd^n x\:(\ln\mathcal{L} + \ln\pi)/2$, such that $\Delta S_\alpha \leq \int\dd^n x\:(\ln\mathcal{L} + \ln\pi)$, bounding the evidence from above, although in most cases this particular bound is diverging.

Finally, expressing the evidence $p(\boldsymbol{D}|M)$ for Gaussian distributions in terms of the R{\'e}nyi-entropy yields
\begin{equation}
p(\boldsymbol{D}|M) = \exp(-\Delta S_\alpha)\sqrt{\mathrm{det}(\boldsymbol{F}+\boldsymbol{P})}\:\left(\frac{2}{\pi}\right)^{n/2},
\end{equation}
if $\delta_\mu = 0$. This relation shows that the evidence is made up from three contributions. On the one hand, it decreases $\propto (2/\pi)^{n/2}$ if the dimensionality of the parameter space, i.e. the model complexity is increased. Also the determinant of the Fisher-matrix generates a scaling $\propto \prod_\mu 1/\sigma_\mu$, such that models with large errors are assigned low evidences as well as being a measure of the dissimilarity of likelihood and prior. Taking the logarithm shows that
\begin{equation}
\ln p(\boldsymbol{D}|M) = -\Delta S_\alpha + \frac{1}{2}\mathrm{tr}\ln(\boldsymbol{F}+\boldsymbol{P}) + n\ln 2 - \frac{n}{2}\ln\left(\frac{2}{\pi}\right),
\end{equation}
meaning that the evidence reflects the inverse volume of the permissible parameter space, allowed by combining likelihood and prior. We emphasise at this point that this compact relationship between evidence $p(D|M)$ and entropy difference $\Delta S$ only exists in the case of tension-free likelihoods, $\delta_\mu = 0$, otherwise an additional factor $1/2$ appears in the exponential term of the Kullback-Leibler divergence which is not present in the evidence.

Comparing Bayes-evidences and information entropy differences shows a clear mathematical relationship between the two, implying perhaps that one could in principle state the relative entropy instead of the Bayesian evidence ratio. The advantage in doing that would be to avoid the empirical Jeffreys-scale. Instead, relative entropies would be stated in units of nats. In addition, the usage of the evidence ratio appears to be motivated by the Neyman-Pearson lemma known from hypothesis testing, where it ensures that two hypotheses are tested against each other with the most efficient test statistic. While it seems to be unclear whether the Neyman-Pearson lemma applies to evidences as well in the sense if evidence ratios constitute the most efficient statistical test to decide between two models, these difficulties are avoided by relative entropies: They are axiomatically defined and unambiguous, and the evidence difference, in units of nats, can be computed for every likelihood, with the only complication originating from having to estimate $\ln p(\boldsymbol{x})$ from samples in the case of non-Gaussian likelihoods or priors. Again, our results will only assume simple forms for Gaussian distributions, and one needs in general to use advanced sampling methods for evaluating evidences in Monte Carlo-Markov chains \citep{mukherjee_nested_2006, kilbinger_bayesian_2010, hou_affine-invariant_2012, lewis_efficient_2013}.

\section{summary}\label{sect_summary}
In this work we have applied information entropy measures to the analysis of cosmological data. Specifically, we computed likelihoods in the Gaussian approximation for spectra of the cosmic microwave background temperature and polarisation anisotropies, CMB-lensing, galaxy clustering and weak gravitational lensing by the cosmic large-scale structure. In almost all application, we work with Gaussian likelihoods and priors and reduce the expressions to be evaluated to operations on Fisher-matrices (or, equivalently, on inverse parameter covariances), for conventional $\Lambda$CDM- and $w$CDM-cosmologies. The motivation of our investigation was to quantify the information content of cosmological probes through probability entropies and show that they can be used for a wide range of applications, not only for making statements about the total error budget but also for quantifying the magnitude of systematical errors and tensions between data sets, as well as Bayesian evidences: The advantage of information entropies is their axiomatic definition and that they quantify in all these applications the extend and location of likelihoods corresponding to different probes with nats as the natural unit, even in the case of model evidences, providing a viable alternative to the Jeffreys-scale.

\begin{enumerate}
\item{We have computed information entropies and demonstrated their decrease through the combination of cosmological probes, and showed how measures of total error derived from the Fisher-matrix, for instance $\mathrm{tr}(\boldsymbol{F})$, $\mathrm{tr}(\boldsymbol{F}^2)$ or $\ln\mathrm{det}(\boldsymbol{F})$ scale with Shannon- and R{\'e}nyi-entropies. Likewise, there are inequalities which bound the relative entropy of the full likelihood through sums of entropies for individual parameters, both marginalised and conditionalised. Values for information entropies, both absolute and relative, are not easy to interpret. For the preferred dimensionless parameterisation used in cosmology consisting of $\Omega_m$, $\sigma_8$, $h$, $n_s$, $\Omega_b$ and possibly $w$, the CMB has the highest information content as evidenced by the most negative numbers for $S$, followed by (tomographic) galaxy clustering, (tomographic) weak lensing and finally CMB-lensing. Differences in entropy are not automatically entropy differences, which makes it difficult to compare numbers for $\Delta S$ with those for $S$: For that reason, we consider probe successive combinations and compute relative entropies relative to the information content of the CMB primaries alone. Kullback-Leibler-divergences $\Delta S$ combine a measure of changed volume of admissible parameter space with the change in the degeneracy directions, and are in that way a measure by how much the degree of knowledge on a parameter set increases. Typical numbers that we have obtained for $\Delta S$ are slightly over one order of magnitude for the full likelihood, although the information entropy of in particular conditionalised likelihoods of individual parameters can change dramatically.}

\item{We looked into three tensions in $\Lambda$CDM- and $w$CDM-cosmologies, the discrepancy in $H_0$ from Cepheid variable stars and the CMB, the tension in $\Omega_m$ and $\sigma_8$ packaged into the parameter $S_8$ between weak lensing by KiDS and the CMB from Planck, and the bias caused by intrinsic alignments in weak gravitational lensing as forecasted for Euclid. In these examples we showed that a quantification of the dissimilarity between the distributions combines the difference in the best fitting values and the width of the distributions, and the difference in their degeneracies in the multivariate cases. Linking relative entropies to traditional quantities in descriptive statistics like $p$-values is possible in the univariate case, but does not show a clear relationship. }

\item{Lastly, we demonstrated the correspondence between information entropy and Bayesian evidence in comparing cosmological models. There is a relation between the entropy difference $\Delta S_\alpha$ and the evidence $p(D|M)$ if $\alpha$ is chosen to be $1/2$, i.e. for the Bhattacharyya-case: This would suggest the possibility of drawing a connection between the Jeffrey-scale and entropy differences measured in nats. Both quantities are clearly related to each other and offer a tradeoff between the goodness-of-fit, as expressed by the ratio of the admissible parameter spaces of two competing models, and a penalty term disfavouring high model complexity $n$.}
\end{enumerate}

One thought that we are currently pursuing is to use the Wasserstein-metric as a generalised symmetric relative entropy for quantifying biases between likelihoods, again with the motivation that relative entropy is axiomatically defined and measured in a natural unit. The Wasserstein-metric overcomes the asymmetry in the definition of relative entropy, making it perhaps a more useful measure of statistical tensions between likelihoods, in particular of Gaussian distributions, where an analytic formula exists.

\section*{Acknowledgements}
AMP gratefully acknowledges the support by the Landesgraduiertenf\"{o}rderung (LGF) grant of the Graduiertenakademie Universit\"{a}t Heidelberg. RR acknowledges funding through the HeiKA-initiative and support by the Israel Science Foundation (grant no. 255/18 and 395/16). AMP and BMS would like to thank the Universidad del Valle in Cali, Colombia, for their hospitality.

\bibliographystyle{mnras}
\bibliography{references}

\bsp
\label{lastpage}
\end{document}